\documentclass[pdflatex,sn-mathphys-ay]{sn-jnl}% Math and Physical Sciences Author Year Reference Style
\usepackage{graphicx}%
\usepackage{multirow}%
\usepackage{amsmath,amssymb,amsfonts}%
\usepackage{amsthm}%
\usepackage{mathrsfs}%
\usepackage[title]{appendix}%
\usepackage{xcolor}%
\usepackage{textcomp}%
\usepackage{manyfoot}%
\usepackage{booktabs}%
\usepackage{algorithm}%
\usepackage{algorithmicx}%
\usepackage{algpseudocode}%
\usepackage{listings}%

\begin{document}

\title[Visualizing Colonial Regimes]{Visualizing Colonial Regimes: A Multi-View Approach to (Historical) Political Transformation}

\author*[1]{\fnm{Nicole} \sur{Husemann}}
\author[1]{\fnm{Steffen} \sur{Kailitz}}
\author[2,3]{\fnm{Christofer} \sur{Meinecke}}

\affil*[1]{\orgname{Hannah Arendt Institute for Totalitarianism Studies}}
\affil[2]{\orgname{Image and Signal Processing Group, Leipzig University}}
\affil[3]{\orgname{ScaDS.AI Dresden/Leipzig, Leipzig University}}


\abstract{This paper takes a critical approach to visualization of colonial regimes. Drawing on critical hermeneutics and postcolonial theory, we created interactive visualizations revealing the constructed nature of colonial categories and exposing the hierarchies between empires and territories. Using the Varieties of Political Regimes dataset, our approach combines temporal flow visualization and geographic distribution mapping. This allows us to contextualize colonial rule within broader patterns of political change. Our visualization challenges conventional, static, and isolated representations of colonial data by making the interpretive frameworks underlying colonial categorization visible through coordinated temporal, spatial, and relational views.}

%%================================%%
%% Sample for structured abstract %%
%%================================%%




\maketitle

\section{Introduction - The Colonial Visualization Gap}\label{sec1}

Colonialism is an essential aspect of modern global history, yet colonial regimes are underrepresented in political science visualizations. Current approaches have several critical limitations. They naturalize violence by presenting colonial boundaries as cartographic facts instead of products of conquest; erase agency by obscuring the indigenous political systems that preceded and resisted colonial rule; simplify temporality by treating colonial periods as discrete units instead of ongoing processes; and hide interpretation by concealing the methodological decisions underlying colonial categorization. The absence of critical approaches to visualization of colonial regimes has significant consequences for how scholars, students, and policymakers understand the relationship between historical colonialism and contemporary global politics. Without visualizations that expose the constructed nature of colonial categories and their persistent influence, analysis of current political formations remains disconnected from their colonial genealogies. This analytical gap becomes particularly problematic when examining questions of political development, conflict patterns, and democratic transitions in formerly colonized territories, where colonial legacies continue to shape political possibilities in ways that conventional datasets render invisible. This paper addresses these limitations by asking what visualization can expose instead of concealing the constructed nature of colonial categories, and how interactive, multi-temporal visualizations might reveal the dynamic relationships between colonial and postcolonial political formations. Drawing from Johanna Drucker’s critical hermeneutics \citep{drucker2011humanities,drucker2014graphesis,drucker2020visualization}, we demonstrate how design choices can challenge rather than reproduce colonial representation. Our visualization contextualizes colonial regimes within 125 years of global political transformation and reveals the constructed nature of regime categories. It also visualizes the temporal dynamics of colonial rule, highlighting patterns of continuity, rupture, and persistence across historical periods. The visualization itself functions as the surface through which the underlying epistemological and classificatory assumptions of regime typologies become visible, interactive, and subject to sustained interpretive engagement. By combining an expandable classification legend, coordinated view-linking, and cartographic infrastructure, the interface creates an interpretive space that supports cross-disciplinary analysis and challenges established political narratives.

\section{Related Work}\label{sec2}
\subsection{Visualization of Colonial Regimes}\label{subsec2-1}

Current approaches to visualizing colonial rule perpetuate the representational limitations that critical scholars aim to overcome. For example, the visualization of Our World in Data of European overseas colonies focuses exclusively on European colonizers and treats colonial relationships through standardized temporal boundaries and categories like ``Colonizers,'' ``Not colonized,'' or ``Multiple colonizers'' \citep{becker2023colonies} Such traditional cartographic representations naturalize sovereignty as bounded territorial entities and fail to capture the contested, fluid nature of colonial and postcolonial political arrangements. Bonilla and Hantel's ``Visualizing Sovereignty'' \cite{bonilla2016visualizing} challenges these limitations by using temporal mapping and animation to reveal the predominance of non-sovereign societies in the Caribbean, rather than treating them as exceptions to normative sovereignty. Their work moves beyond static geographic positioning, showing how societies shift between different political statuses over time. This reveals the ``uneven and nonlinear process'' of decolonization and calls for ``prophetic cartographies'' that can imagine alternative political futures. However, their regional focus on Caribbean non-sovereignty, while methodologically innovative, does not address the global patterns of colonial regime persistence and transformation across the full 125-year scope of modern colonialism examined by our approach.

\subsection{Visualization Practices in Political Science}\label{subsec2-2}

Visualization practices in political science, as surveyed by \cite{han2024survey} and \cite{traunmuller2020visualizing}, have expanded, but remain shaped by positivist assumptions. Visual representations such as maps, timelines, and scatterplots are widely used to present regime classifications as objective facts with limited critical engagement. However, the survey by \cite{han2024survey} primarily considers visualization techniques published in visualization research venues while characterizing political science visualization practice based on only three general-interest U.S. journals. Important political science projects such as V-Dem \citep{vdem2025}, whose visualizations, though not innovative from a visualization research perspective, are vital for political analysis, are overlooked. This exclusion reflects a broader limitation, as it disregards internally developed visualization practices within political science itself. Moreover, the inclusion of American Politics as a primary keyword introduces a thematic bias, favoring U.S.-centric political structures and sidelining phenomena such as multiparty systems and authoritarian regimes, thereby narrowing the survey’s relevance to global regime research.

\begin{figure}[tb]
\centering
\includegraphics[width=0.85\textwidth]{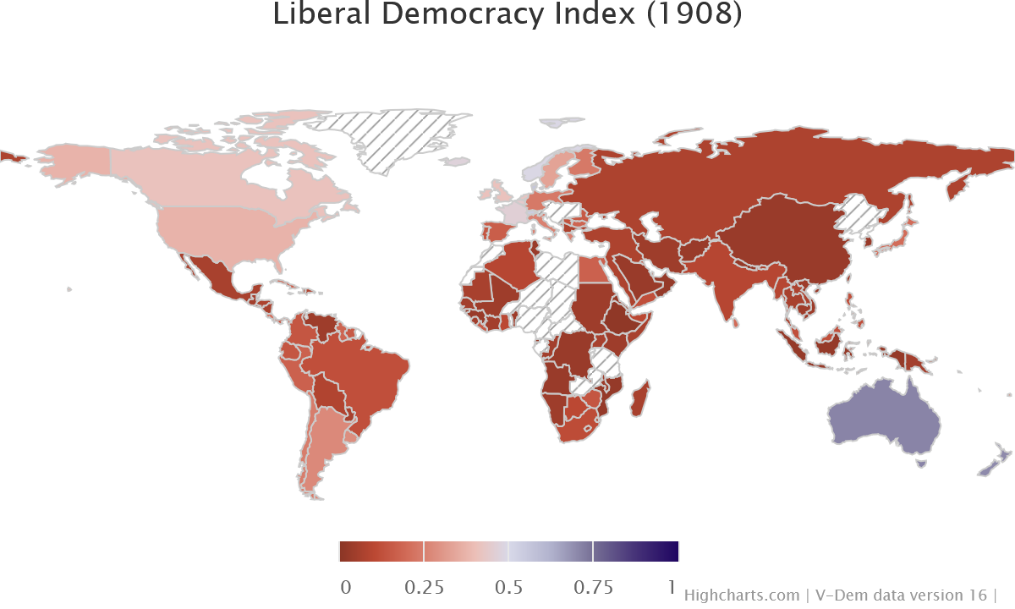}
\caption{V-Dem Liberal Democracy Index, 1908 (V-Dem v16).}\label{fig1}
\end{figure}

V-Dem's visualizations illustrate the interconnection of dataset choices and the resulting visualization consequences in two distinct ways. First, their classification, which uses institutionally based indices, flattens lack of democratic rights into a numeric scale, irrespective of whether the regime is externally enforced or not, an appropriate design choice for comparing sovereign states but one that renders colonies equivalent to other types of closed autocracies such as autocratic monarchies or communist ideocracies. Moreover, their dataset is built around formally independent states, a scope that serves their comparative purpose, but that also excludes colonial territories. Both consequences are directly visible in V-Dem's choropleth map in Figure~\ref{fig1}.
Viewers preattentively register colonized Eastern Africa and Western Asia's Autocratic Monarchies as equivalent: in the 1908 Liberal Democracy Index, both occupy nearly the same position on the diverging color scale, collapsing externally imposed colonial exclusion and domestic autocratic rule into a single undifferentiated category, while large areas of Middle Africa appear as hatched white gaps, without any information or explanation.
In both cases, the map's own visual grammar obscures the political judgment embedded in its construction: substantively, by treating colonial domination as equivalent to other forms of unfreedom; cartographically, by rendering its absence from the dataset as something that simply is given, rather than something that was made.

\subsection{Johanna Drucker's Approach to Visualization}\label{subsec2-3}

Nothing, however, is simply given. Drucker challenges the assumption that data are objective givens by introducing the term capta to emphasize their constructed nature. Her concept of probabilistic critical hermeneutics provides valuable guidance for visualizing complex historical processes like colonialism. It reframes visualization as a fundamentally interpretive practice shaped by context, history, and uncertainty.

Visualizations are a situated act of knowledge production that organizes meaning through design choices in scale, color, and spatial arrangement. Drawing on hermeneutic and postphenomenological thought, Drucker argues for transparency of interpretation, contextual embedding, and interpretative openness. Her approach treats visualizations as dynamic, participatory interventions instead of static representations, insisting that they must foreground their own epistemological premises and enable critical engagement with the structures they depict \cite{drucker2011humanities,drucker2013performative,drucker2014graphesis,drucker2020visualization}. These concerns align with broader digital humanities critiques of data neutrality \citep{dignazio2023data,risam2018new} and approaches that transform archival absences into analytical opportunities through visualization. \cite{dork2013critical} and \cite{klein2013image} advance a critical approach to information visualization that stresses disclosure, contingency, and plurality. Critical visualization practices reject the assumption of neutrality, recognizing the value-laden choices embedded in data selection, encoding, and presentation. Visualization becomes a political act, shaping what is seen and what remains invisible. Similar objections surface in digital humanities scholarship that interrogates the apparent neutrality of computational methods \citep{liu2012cultural,mcpherson2012digital} and examine how digital representations can perpetuate historical inequalities \citep{noble2018algorithms}. Michael \cite{10.1145/3290605.3300418} names this directly as an ethical demand rather than only a critical observation: ``we have an obligation, where possible, to make these invisible facets and contributions visible.''  What Drucker and Dörk et al. describe as a political effect of visualization, Correll recasts as a responsibility incumbent on those who design it.

\subsection{Postcolonial Critique and Epistemic Violence}\label{subsec2-4}

Visualizations of Colonial rule were never merely neutral depictions of territorial order. They actively contributed to justifying and stabilizing it. As Edward \cite{said1978orientalism} shows in \emph{Orientalism}, colonial knowledge was based on representations that portrayed colonized spaces as backward, chaotic, or dangerous---and thus controllable. Maps, diagrams, and statistical classifications played a central role in this: they made colonial boundaries and categories appear to be natural conditions.

Charles \cite{mills2007white} extends this account by theorizing such exclusions as a structurally sustained epistemology of ignorance: not a passive absence of knowledge but an actively produced and socially maintained not-knowing, in which dominant groups develop a positive cognitive investment in not perceiving certain social and political realities, particularly those bound up with race and colonial domination. On this view, the absence of colonized populations from political knowledge production is actively reproduced by the same categories, instruments, and institutions that determine what counts as knowable in the first place; the categories themselves perform the occlusion, rather than failing to capture a reality that exists independently of them.

Mills's account, read alongside Said's, implies a colonial invisibility that is layered, locatable, and reversible, produced by specific classificatory, relational, and cartographic mechanisms; \cite{santos2007beyond} notion of abyssal thinking locates comparable exclusions beyond any possibility of interpretive recovery, a line more absolute than the one Mills describes.

In the Varieties of Political Regimes classification used in this work, colonial regimes are explicitly coded---especially in the case of direct colonial rule---as among the most coercive forms of closed autocracy \citep{kailitz2026vaporegcodebook}. These regimes typically eliminate political participation, suppress civil liberties, and impose external authority without consent or accountability. The colonial executive, appointed by and accountable only to the metropole, operates entirely beyond local constraints, bypassing any form of domestic checks or representative legitimacy \citep{mamdani2018citizen}. Crucially, direct colonial regimes were almost invariably structured along rigid racial hierarchies, institutionalizing systemic exclusion and legitimizing violence as an instrument of governance \citep{fanon1968wretched,young1995african}.

\section{Varieties of Political Regimes Dataset}\label{sec3}

This section describes the Varieties of Political Regimes (Va-PoReg) dataset Version 3.2 \citep{kailitz2026vaporeg}, developed at the Hannah Arendt Institute for Totalitarianism Studies, which provides the basis of our work. It explains how the dataset defines its units of observation, classifies political regimes, and incorporates colonial and non-sovereign political entities. Va-PoReg systematically codes political regimes across all states and selected non-sovereign entities from 1900 to 2025, encompassing over 29,000 country-year observations. The unit of observation is a political entity in a given year. Unlike other regime datasets that restrict their coverage to formally sovereign states, Va-PoReg also includes colonial and dependent territories as political units in their own right. This allows political regimes to be coded for territories that did not possess formal sovereignty instead of excluding them from the analysis or assigning them the regime characteristics of a later sovereign successor.
 
Va-PoReg's defining feature lies in the way these political entities are classified: regime types are distinguished according to their principal sources of legitimation and their structural configurations \citep{kailitz2009varianten,Kailitz01012013,kailitz2026vaporegcodebook}. The former refers to the principles through which political authority is justified, while the latter captures the institutional structures through which it is exercised. Accordingly, Va-PoReg distinguishes 22 regime types based on their principal sources of legitimation, such as ideological, monarchic, personalist, or religious claims to authority. These categories are complemented by four structural dimensions that capture the institutional organization of political authority: participation, competition, executive constraints, and political liberties. Participation captures de facto suffrage and electoral inclusion, while competition concerns party pluralism and the meaningfulness of elections. Executive constraints capture limitations on executive power, and political liberties refer to freedoms of expression, assembly, and organization. This dual approach captures both the justificatory principles and institutional architecture of political regimes across time and space. Sovereignty constitutes a separate variable in Va-PoReg, coded as sovereign, semi-sovereign, or non-sovereign, orthogonal to a unit's legitimation pattern: a political unit can be non-sovereign and still carry a full regime classification. Direct and indirect colonial rule are not gradations of sovereignty; both remain non-sovereign as long as domain control and override capacity rest with the metropole. A unit is coded under indirect rule only where three conditions jointly hold: indigenous institutions exercise limited but genuine authority over defined domestic policy domains; this authority derives from a source of legitimation independent of colonial appointment; and the metropole retains domain control over core areas such as economic, monetary, or judicial policy alongside override capacity over delegated domains, evidenced by reserve powers, gubernatorial veto rights, or the power to dissolve indigenous legislatures or dismiss ministers. Direct rule applies where such an independent, binding source of indigenous authority is absent. A unit crosses into semi-sovereignty only where two conditions hold simultaneously: external control is confined to foreign policy and defense, and the external power holds no veto, suspension, or dissolution power over indigenous institutions within their domestic domains; either condition alone leaves a unit within the colonial family \citep{kailitz2026vaporegcodebook}.
 
For each entity-year combination, multiple regime types are assigned based on different classification schemes: a detailed classification of 22 legitimation-based regime types, a compact classification aggregating conceptually related types, and simplified typologies including quadruple, triple, and binary classifications that facilitate broad cross-national and longitudinal comparisons \citep{kailitz2026vaporegcodebook}. While the quadruple classification is structurally similar to \cite{luhrmann2018regimes}'s Regimes of the World (RoW) typology, Va-PoReg deliberately draws 
boundaries between categories differently to reflect its conceptual distinctions and, unlike RoW, is not restricted to currently sovereign states. Regional assignment follows the United Nations Statistical Division scheme, enabling allocation to standardized global regions. The dataset enables diachronic analyses of regime change and visualization of temporal patterns and typological transformations.

\section{Visualization Design}\label{sec4}

We visualize political regimes with an Alluvial Diagram showing regimes from 1900 up to 2025 as flows through time and a geographical view that highlights the spatial distribution of a selected regime type. 
The interactive web application with all visualizations is available at: \url{https://www.va-poreg.de/alluvial_countries}, and we encourage direct exploration.
The coordinated interaction design makes visible the historical transformations and continuities often obscured in static political charts. The following subsections first outline the analytical tasks the visualization supports, then turn to the three coordinated components that translate Va-PoReg's classificatory decisions into interactive surfaces open to critical interrogation.
Section~\ref{sec5} then walks through these three components in the order a user would traverse them, using France's regime trajectory from 1900 to 2025 as the intended path through the coordinated views. 

\subsection{Users, Analytical Goals, and Visualization Tasks}\label{subsec4-0}

The visualization is designed for scholars interested in the
historical transformation of political regimes and the role of
colonial rule in shaping these transformations. Its primary users
include political scientists studying regime change and persistence,
scholars of colonial and postcolonial studies examining imperial
relationships and processes of decolonization, and area studies
scholars interested in regional trajectories of political change.
While these communities approach the data from different perspectives,
they share a need to examine regime classifications not only as
descriptive categories but also as analytical constructions that shape
which historical patterns become visible.

To support these perspectives, we identify three interconnected groups
of analytical tasks: temporal tasks, spatial and relational tasks, and
comparative and interpretive tasks. These tasks provide the analytical
rationale for the coordinated views and their interactive mechanisms.

\textbf{Temporal tasks} involve tracing regime trajectories across the
1900--2025 period, identifying periods of regime stability and
transition, and examining the timing and direction of political
transformations. Users may, for example, follow the persistence of a
particular regime type across several periods, identify periods in
which transitions were particularly frequent, or examine how colonial
regimes relate temporally to processes of democratization,
decolonization, and authoritarian persistence. The alluvial view
supports these tasks by representing regime types and transitions
across user-selected time points and allowing users to follow flows
between successive periods.

\textbf{Spatial and relational tasks} involve examining both the
geographic distribution of regime types and the relationships between
colonial empires and the territories under their control. Users can
investigate where particular regime types were concentrated, compare
regional patterns across different historical periods, and examine the
geographic scope of individual colonial empires. For colonial regime
types, users can further identify which territories were associated
with a particular empire and compare the number and distribution of
colonies across imperial powers. The geographic view supports these
tasks by linking a selected temporal point and regime type to its
spatial distribution and by making empire--colony relationships visible
through coordinated highlighting.

\textbf{Comparative and interpretive tasks} involve examining how
classificatory and methodological decisions affect the patterns that
become visible in the visualization. Users can compare aggregated and
differentiated regime classifications, investigate how colonial
regimes are represented within broader regime categories, and examine
how the inclusion or exclusion of sub-regimes changes apparent
patterns of transition and persistence. These comparative operations
also support an interpretive task: questioning the categories
themselves and reflecting on how different ways of classifying
political regimes produce different historical representations. Rather
than treating regime type classifications as neutral and given, users
can explore what becomes visible or obscured when categories are
aggregated or differentiated.

The remainder of this section presents the visualization's three coordinated components. We first describe how classification choices are made explicit and interactively adjustable (Section~\ref{subsec4-1}), before turning to the alluvial diagram that renders regime trajectories over time (Section~\ref{subsec4-2}) and the geographic view that situates them in space (Section~\ref{subsec4-3}). Classification is addressed first because the choices it exposes shape how the temporal and spatial views can subsequently be read.

\subsection{Making the Constructedness of Colonial Regime Classifications Explicit}\label{subsec4-1}

\begin{figure}[htb]
\centering
\begin{minipage}[b]{0.48\textwidth}
\centering
\includegraphics[height=4.5cm]{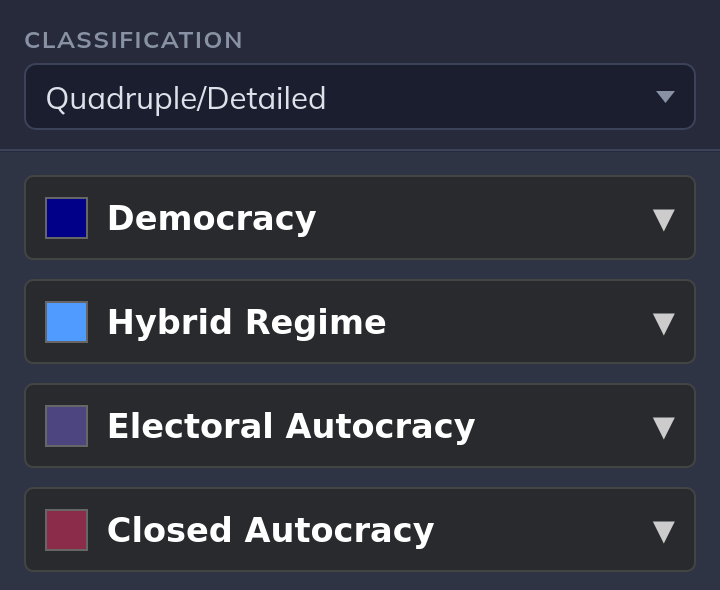}\\[2pt]
{\footnotesize (a)}
\end{minipage}
\hfill
\begin{minipage}[b]{0.48\textwidth}
\centering
\includegraphics[height=4.5cm]{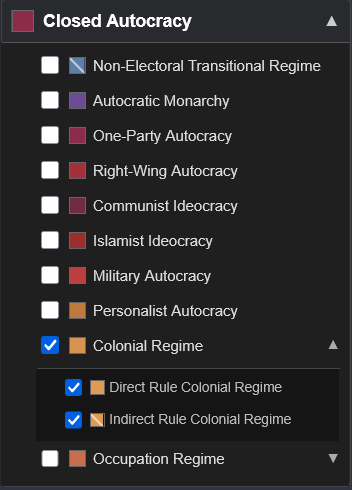}\\[2pt]
{\footnotesize (b)}
\end{minipage}
\caption{The legend before (a) and after (b) interaction. Expanding Closed Autocracy exemplifies how the quadruple classification aggregates several distinct legitimation-based regime types.}\label{fig2}
\end{figure}

Our visualizations directly challenge and disrupt the naturalization of colonial categories by making their interpretive foundations visible through progressive disclosure and coordinated interaction. The interface initially presents Va-PoReg's quadruple classification system (see Figure~\ref{fig2}(a)) with visual cues that intuitively suggest the possibility of expanding these categories to show a more differentiated regime types, including Colonial Regime, each allowing further expansion to distinguish between direct and indirect rule (see Figure~\ref{fig2}(b)). This expandability is where the legend enacts Drucker's notion of capta: presenting the regime categories not as fixed labels but as expandable and recombinable shows them as interpretive constructions and thereby exposes the constructed nature of all data categorizations, transforming what might appear as neutral regime types into explicitly interpretive frameworks that enable particular kinds of analysis while foreclosing others \citep{drucker2014graphesis}. Tooltips accompanying each regime type provide brief definitions, supporting this exploration with conceptual transparency. The toggle between different classification schemes demonstrates how methodological decisions directly influence visual patterns. When users expand the initial categories to explore more detailed classifications, they encounter the underlying complexities that are obscured by the aggregated quadruple, triple, and binary typologies. By coordinating this classificatory complexity across the temporal, spatial, and relational views, users can trace how these different colonial strategies shaped distinct patterns of regime transition and persistence. 

\subsection{Temporal Flow Visualization of Regime Changes}\label{subsec4-2}

\begin{figure}[tb]
\centering
\includegraphics[width=0.85\textwidth]{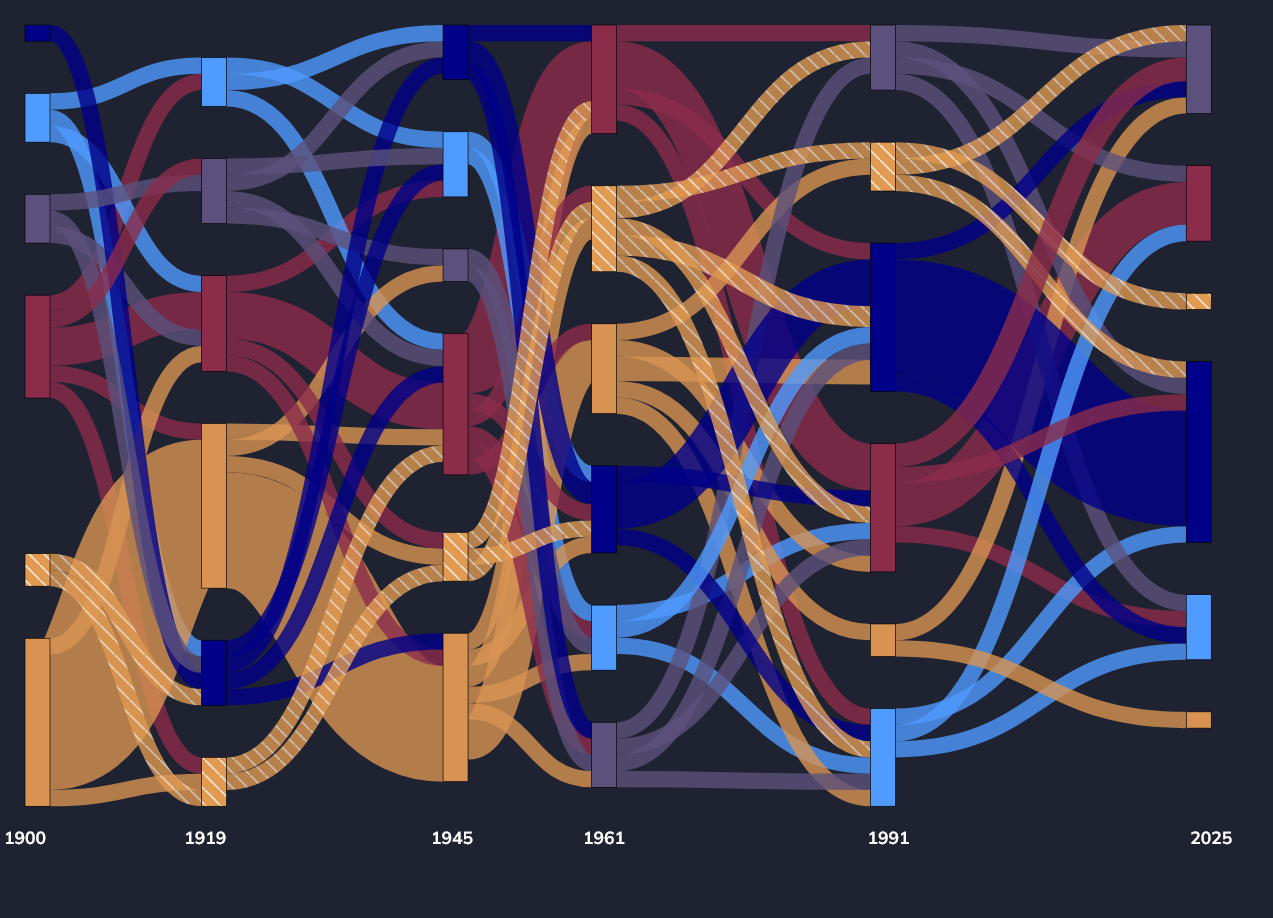}
\caption{Alluvial Chart with regime transitions between Democracy, Hybrid Regime, Electoral Autocracy, Closed Autocracy Direct and Indirect Colonial Regime.}\label{fig3}
\end{figure}

In order to show regime transitions over time, we visualize them with an Alluvial diagram, which can be seen in Figure~\ref{fig3}. The x-axis is divided into six temporal segments, and the y-axis is used to space out the nodes of the diagram.

Initially, the nodes are set at six data-derived years: 1900 and 2025, marking the dataset's temporal span, and 1919, 1945, 1961 and 1991, computationally identified as peaks of regime change across the dataset. Users are not bound to this default segmentation, however: the control panel allows these years to be freely added or removed, extending the same interpretive agency afforded by the expandable legend (Section~\ref{subsec4-1}) to the temporal axis itself. Each node represents a regime type in a temporal segment and aggregates all countries that held that regime in the specific year corresponding to the node. The width of a node indicates the number of countries with each regime in a given temporal segment. The edges between the nodes are Bezier curves that represent transitions between regime types, and their width indicates the number of transitions between two regime types. This allows a direct comparison of transition volumes between different regime categories. The nodes and edges are colored based on the regime type. The colors depend on the selected regime classification scheme, which can be changed by the user but are consistent between them. In addition, sub-regime types can be added or removed. In the case of adding sub-regimes, they are excluded from the parent category and treated as their own regime type. An analysis of regime durability is also possible by observing persistent nodes with strong self-connections across periods indicating stable regime types. When hovering over a node, the number of countries with that type of regime is displayed. Additionally, all edges and nodes are faded out except those that belong to earlier time steps and eventually lead to the selected node, highlighting the transitions that culminate in the current regime type. 
%An example can be seen in Figure~\ref{fig6}.
By hovering over a link, the regime transition is shown with the number of countries.

Further, to get more details for a regime node, a user can click on it to see the Geographical view (see Subsection~\ref{subsec4-3}) with information about the empire-colony relation. In general, users can follow specific regime types through time by using hover highlighting, observing how they emerge, grow, shrink, or disappear.

\subsection{Geographic Distribution of Regime Types}\label{subsec4-3}

\begin{figure}[tb]
\centering
\includegraphics[width=0.85\textwidth]{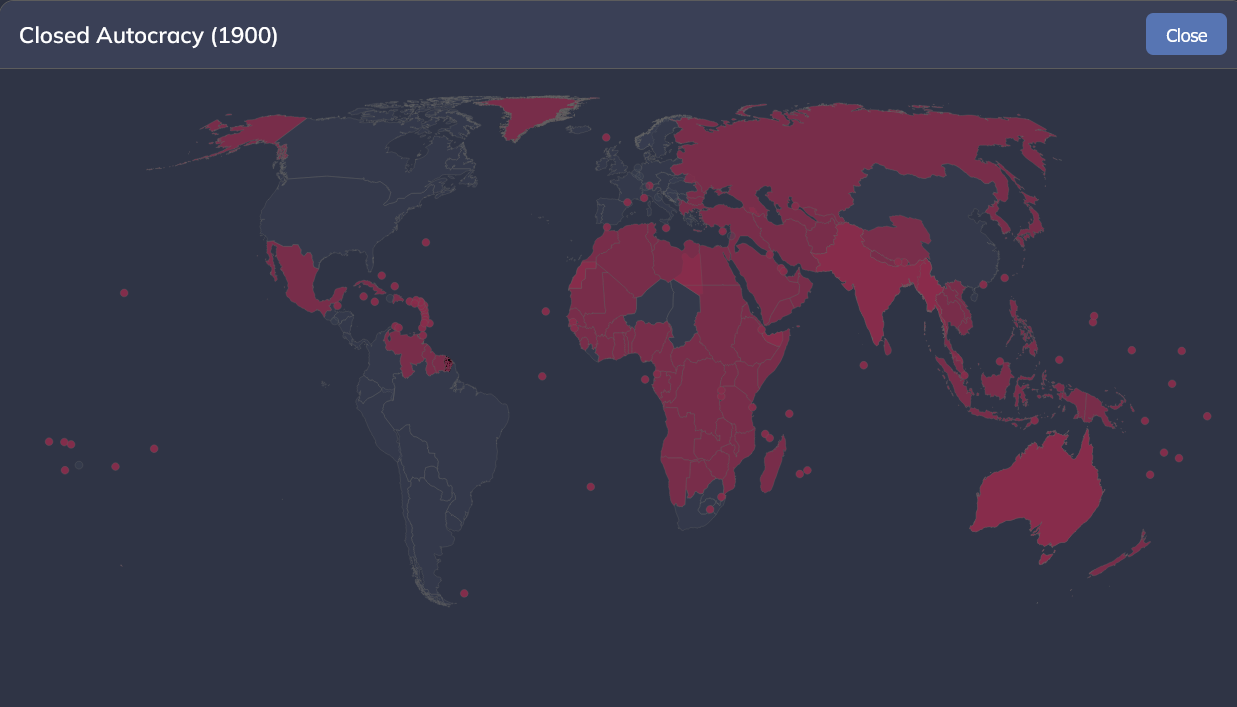}
\caption{Cartographic view of Closed Autocracy Node 1900.}\label{fig4}
\end{figure}

The geographical view shows a Choropleth map of countries with the selected regime type using the Equal Earth projection \citep{vsavrivc2019equal}, which retains the relative size of countries. By using this projection instead of a conventional like Mercator's, this map avoids the cartographic distortions that historically reinforced Western dominance by making colonial Empires appear larger than their colonies \citep{hirsch2020postcolonial}. The map uses a custom made TopoJSON, built on CShapes \citep{schvitz2022mapping} for its representation of border changes over time and supplemented with individually sourced GeoJSON geometries for the polities Va-PoReg covers but standard geographic data sources omit. Territories whose bounding-box area falls below a fixed pixel threshold are rendered as small circular markers at their geographic centroid instead of as polygons, keeping units Va-PoReg codes as regime observations visible even where their geometry would otherwise be illegible (see Figure~\ref{fig4}).

The view also shows the selected year and a list of all countries with the selected regime type. This provides spatial context while maintaining temporal specificity. For colonial regime nodes, the geographical view further includes a list of empires with the number of colonies per empire showing major colonial powers. Furthermore, by hovering over a specific colonial empire, all empire-colony relationships are highlighted by a white border.

\section{Nested Occlusion: The Case of France, 1900--2025}\label{sec5}

The limitations of conventional visualizations of regime types can be traced through a single empirical thread running the full temporal length of the Va-PoReg dataset: France, from 1900 to the present day. The exclusion problem is already resolved at the level of the dataset, since Va-PoReg codes non-sovereign colonial territories as first-class regime observations (Section~\ref{sec3}); this case study demonstrates a distinct problem, occlusion, in which data already present in the dataset remains concealed through classification, relational structure, and cartographic representation. Each of these three layers pairs a specific analytical problem with a Va-PoReg decision that makes it addressable and an interface mechanism that renders that decision visible and navigable; underlying all three is the coding of sovereignty status independently of regime type, the condition that allows non-sovereign colonial territories to enter the dataset's scope as regime observations in the first place.

\subsection{First Layer: Classification Occlusion}\label{subsec5-1}

The alluvial visualization is first set to the quadruple classification, which is directly comparable to the standard typology used by V-Dem and similar datasets: democracy, hybrid regime, electoral autocracy, closed autocracy. This initial view presents an aggregated flow of regime transitions. In 1900, France appears within the Hybrid Regime stream, classified at the detailed level as an Electoral Oligarchy. Its 27 colonial territories are absorbed, without distinction, into the Closed Autocracy node.

The legend is the first instance of the anti-occlusion intervention. The interface initially presents the quadruple classification, but visual affordance indicates that these categories are not fixed: a  classification-scheme selector at the top of the legend, labeled 'Quadruple/Detailed' (Figure~\ref{fig2}(a)), and an expand chevron beside each category. Expanding Closed Autocracy reveals ten sub-types, including Colonial Regime. Selecting Colonial Regime causes a new node and stream to detach from the Closed Autocracy flow and render its node and stream across the temporal axis. The resulting significant shrinkage of the Closed Autocracy node as the colonial stream separates immediately makes the occlusion legible.

The legend functions therefore as an argument in itself: collapsing or differentiating regime types is a methodological decision with political consequences, and the interface makes this visible and reversible.

\subsection{Second Layer: Relational Occlusion}\label{subsec5-2}

With the Colonial Regime stream now visible, the alluvial exposes an embedded structure: the large colonial flow behaves categorically different from every other regime type. It does not gradually democratize, or transition through hybrid forms. Instead, it persists with unusual stability across the first half of the twentieth century and then collapses abruptly, dispersing in various other regime types. A thin but unbroken thread continues to the present, directly contradicting the common view that colonization is a past phenomenon.

\begin{figure}[tb]
\centering
\includegraphics[width=0.85\textwidth]{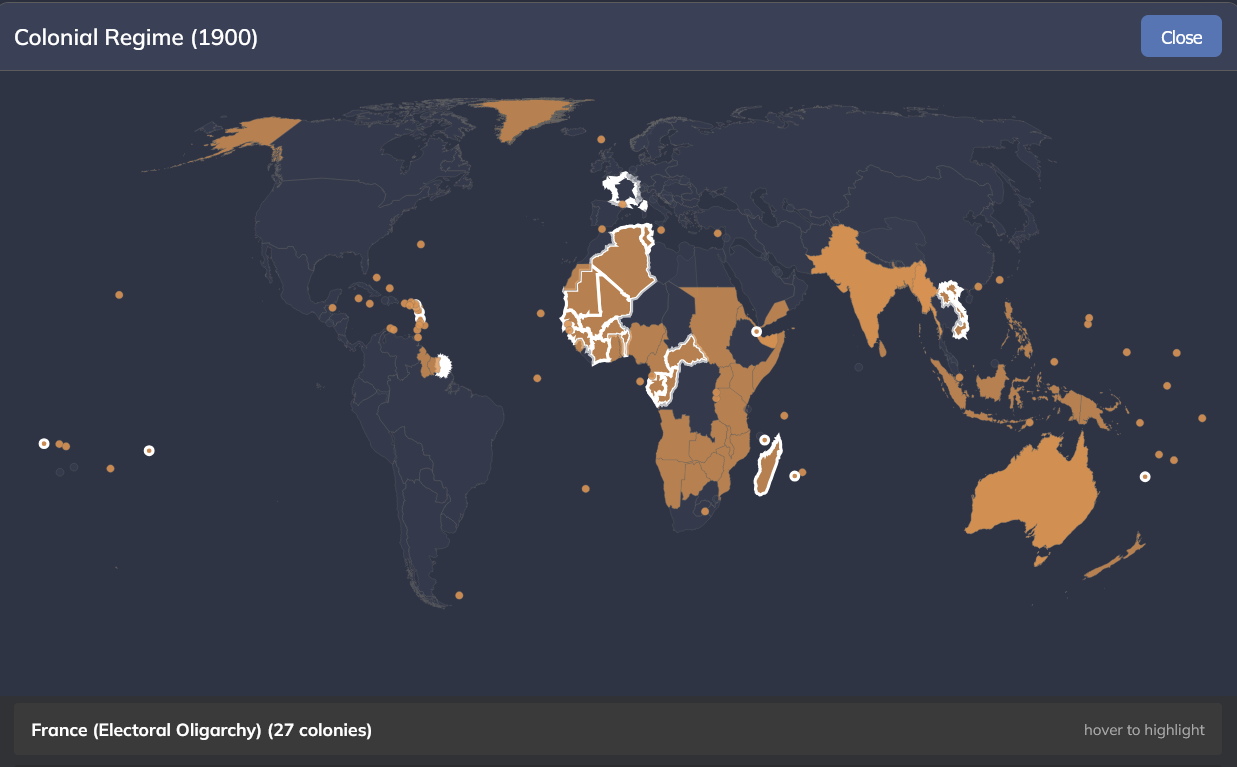}
\caption{Cartographic view of Colonial Regime Node 1900. France and governing territories are highlighted in white.}\label{fig5}
\end{figure}

Clicking the Colonial Regime node opens the geographic view and resolves the flow's anonymous aggregation. In 1900, France appears as a colonial empire governing 27 territories across Africa, Asia, and the Pacific (see Figure~\ref{fig5}). By 1945, France is coded as a Liberal Democracy. It has emerged from occupation by Nazi Germany and rebuilt on a democratic constitutional foundation, yet clicking the 1945 Colonial Regime node reveals it simultaneously governing 28 colonial territories under colonial rule, denying their populations any political participation, civil liberties, or recourse against an executive only accountable to France. The interface places these facts in the same typological space. The contradiction is not immediately discoverable, instead, the user must navigate to it.

Thus, the second occlusion is not just that colonial territories are absorbed into a broader category, but that the relationship between the colonial states and the democratic metropole is invisible until actively surfaced through interaction.

\subsection{Third Layer: Cartographic Occlusion and Temporal Persistence}\label{subsec5-3}

The geographical view reveals a third layer of occlusion operating below classification and relational structure: several French colonial territories in the nodes from 1900 until 1960 fall below the visibility threshold introduced in Section~\ref{subsec4-3}, and appear instead as circular markers, making legible what would otherwise vanish from the visualization.

\begin{figure}[tb]
\centering
\includegraphics[width=0.85\textwidth]{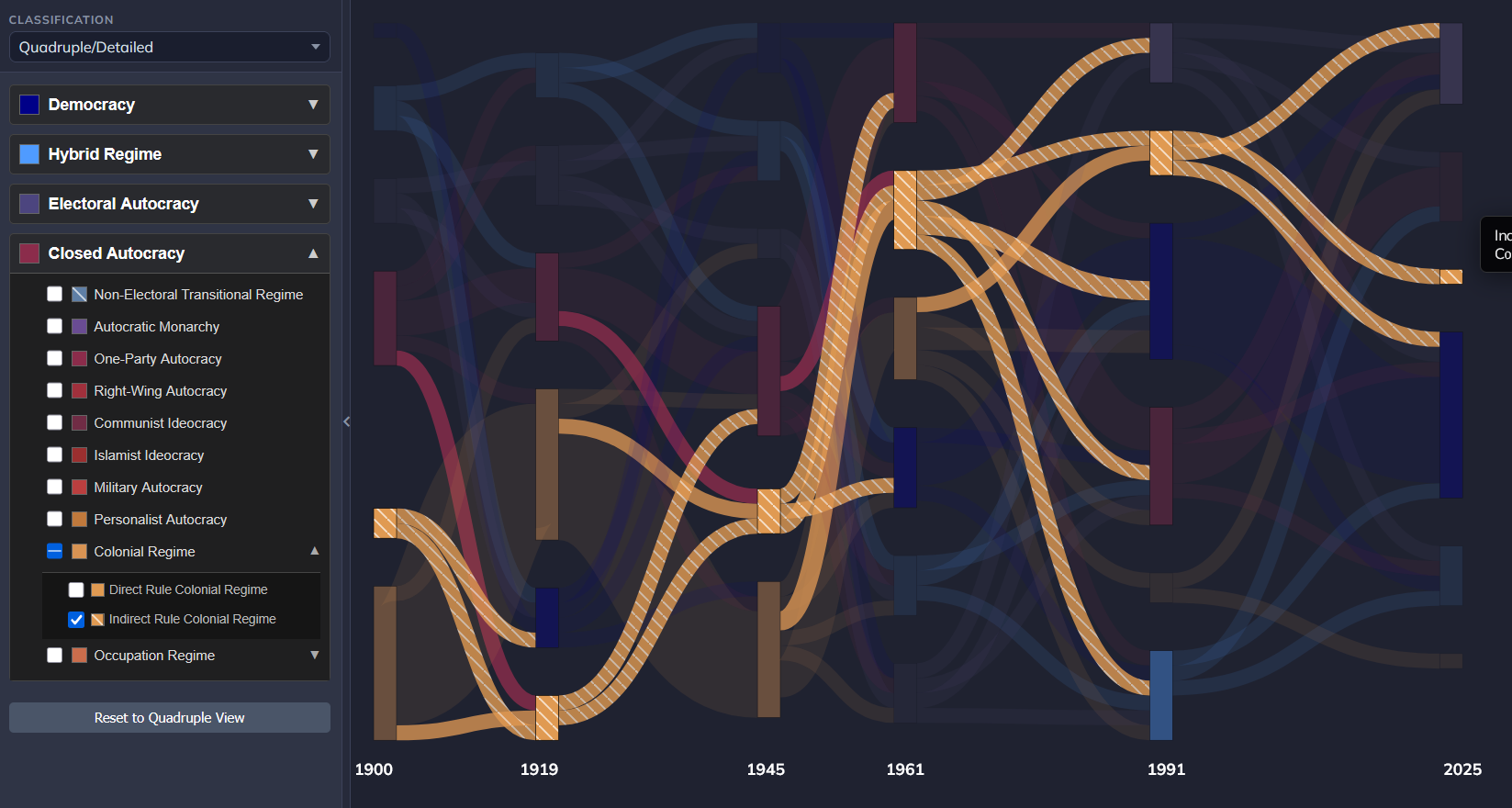}
\caption{Highlighting of Indirect Rule Colonial Regime Trajectory while hovering over its 2025 Node.}\label{fig6}
\end{figure}

The France thread continues beyond 1960 (see Figure~\ref{fig6}). Va-PoReg codes French Polynesia as an Indirect Rule Colonial Regime in 2025, creating tension with how conventional datasets handle oversea territories, which are typically coded as integral parts of the metropolitan state. The persistence of this indirect relationship across more than a century, across the formal moment of decolonization, demonstrates that colonial structures are not historical episodes but ongoing configurations. The conventional periodization that treats the 1960s as the end of colonialism is itself a product of the same classificatory choices that the quadruple scheme embeds and our visualization seeks to make visible.

Taken together, the France thread demonstrates that occlusion in the visualization of  colonial regimes is not a single problem with a single solution. It is layered: the classification system hides colonial territories within a broader autocracy category; the default relational view hides the colonizer-colony relationship until the user actively reveals it; and the cartographic infrastructure hides the smallest territories entirely. Our visualization intervenes at each layer independently, and the interventions compound. It is only when all three are addressed simultaneously that France in 1945 becomes fully legible as a political formation: a Liberal Democracy governing 28 colonial territories, some too small to see on regular cartographic applications.

\section{Critical Visualization Contributions}\label{sec6}

Our approach operationalizes critical hermeneutics by preventing the naturalization of colonial categories through progressive disclosure of interpretive complexity. Our expandable classification system replaces a homogeneous presentation of colonial regimes with a differentiated one, exposing the constructed nature of colonial typologies. When users expand the ``Colonial Regime'' entry via its disclosure triangle in the legend to reveal direct and indirect rule distinctions (Figure~\ref{fig2}(b); terminology per Section~\ref{sec3}), they encounter the analytical consequences of different classificatory choices, making visible the methodological decisions that conventional visualizations conceal. Following \cite{mills2007white}, the absence of colonial territories from conventional regime datasets is better understood as a structurally sustained epistemology of ignorance than as a neutral gap in data collection: the classification schemes themselves perform the occlusion; they do not merely fail to record a reality that exists independently of them. Read this way, the disclosure mechanism does more than let users explore subcategories --- it reverses the specific classificatory choice that caused colonial regimes to disappear into ``Closed Autocracy'' in the first place. This is only possible because such occlusion is produced by nameable mechanisms, not by an unbridgeable epistemic line.

This antinaturalization strategy functions through coordination between the temporal and geographic views, preventing the isolation of colonial data by contextualizing colonial territories within their persistence over time, geographic clustering, and hierarchical relationships with imperial powers. It operationalizes what \cite{dork2013critical} term disclosure and plurality: distributing a single argument across temporal, spatial, and relational aspects gives users multiple independent entry points into the same classificatory complexity, rather than a single vantage point that a viewer could take at face value. Functioning as a boundary object~\citep{star1989institutional}, the interface is designed to support exchange between epistemic communities, such as regime analysts, postcolonial theorists, and digital humanists, without enforcing consensus. Instead, it is meant to create space in which disagreement can function as an analytic resource. Following Drucker's emphasis on historical situatedness, examining colonial regimes requires considering their broader patterns of political transformation. The France case (Section~\ref{sec5}) illustrates the interpretive tension this produces: a state governing itself democratically at home while simultaneously ruling dozens of territories under authoritarian rule abroad. This contradiction reflects a deeper paradox at the heart of modern democratization: key historical drivers like France and the United Kingdom maintained the world's largest colonial empires well into the late twentieth century. Even as both states expanded electoral participation and civil liberties at home, they imposed authoritarian rule abroad through colonial governance that denied representation or rights to colonized populations, routinely relying on coercion, surveillance, and institutionalized racial hierarchies. While forms of colonial violence changed and in some contexts diminished over time, their structural imprint persisted in administrative practices and legal exclusions that endured long after formal decolonization. This tension highlights the inadequacy of regime typologies restricted to metropolitan contexts. This also enacts \cite{dignazio2023data} principle to examine power: rendering the empire-colony hierarchy not as an implicit backdrop to the map but as an explorable relation exposes precisely the structural asymmetry between metropole and territory that conventional choropleths leave unmarked. By embedding colonial regimes into the same typological space as sovereign states, our visualization exposes this asymmetry and challenges the notion of internally coherent democratic trajectories. Instead of presenting decolonization as a finite historical episode, our visualization demonstrates the endurance of colonial structures decades after formal independence. The persistent influence of colonial regimes during the post-Cold War era and beyond challenges the conventional periodization that considers the 1960s as the ``end'' of colonialism. Most notably, our visualization reveals that indirect colonial relationships persist into 2025 - there are still indirect colonies today. The persistence of colonial relationships, rendered invisible in conventional political datasets that code these territories simply as ``independent states,'' demonstrates the potential of visualization to challenge naturalized assumptions about political sovereignty and historical periodization. These findings do not speak for themselves. Their interpretation, and the assumptions they challenge, depend on how colonialism is visualized, categorized, and made analytically accessible. This is where the epistemological design of our visualization becomes central. Our visualization is not a neutral interface but a reflective medium for producing knowledge that highlights interpretive diversity.
This reflective character is itself an ethical commitment, not only an analytical one: \cite{10.1145/3290605.3300418} argues that visualization researchers have a duty to make invisible labor, uncertainty, and impact visible wherever possible.

\section{Limitations \& Future Work}\label{sec7}

Many limitations are based on the underlying data. For example, our and most other regime visualizations approaches do not capture sub-national (colonial) relationships, which limits analysis to the state level. This ignores ethnic or regional conflicts influenced by colonial occupation as different regions within the same colony experienced different levels of control. The Alluvial diagram faces certain challenges related to political reality. For example, entities like Transvaal, Natal and Cape Colony only existed until 1909, after which they became part of South Africa, potentially making the total number of transitions seem misleading. Although our approach offers several regime classifications, the simplification into specific regimes for visualization may obscure important complex relations, or distinctions between different classifications. Selecting only a subset of classes may help a user navigate the visualization better but could also lead to the loss of information. In addition, our approach only focuses on formal political control, which ignores forms of economic or cultural colonialism. Therefore, the visualization might miss ongoing influence, as these forms of economic and cultural domination constitute what scholars term the ``coloniality of power'' \citep{quijano2000coloniality} and ongoing neo-colonial arrangements \citep{young2001postcolonialism}. 
Beyond these data-related limitations, the visualization has not yet been evaluated with its intended user groups, though informal feedback from social science collaborators on the Va-PoReg project helped identify interface issues and clarify interaction mechanisms along the way.
The task analysis presented in Section~\ref{subsec4-0} was derived analytically from the literature and from our own engagement with the Va-PoReg dataset.
A user study involving domain experts is planned to examine if the coordinated views indeed support the temporal, spatial and relational, and comparative and interpretive tasks the analysis calls for. It will also assess whether the progressive disclosure of classification decisions offers a suitable framework for critical reflection on the regime categories and is not easily overlooked or simply treated as a technical filtering option. This would also help identify usability barriers, particularly for users less familiar with interactive visualizations, and could surface additional tasks not anticipated in the current design.
In future work, a focus on visualization methods that could show the ongoing colonial legacy could provide a broader picture by including economic and cultural indicators. This would also connect the past to the present.

\section{Conclusion}\label{sec8}

This paper shows that critical visualization design can expose rather than conceal colonial power structures through three specific interventions: progressive disclosure of Va-PoReg's classificatory complexity, coordinated linking between temporal and geographic views, and a cartographic infrastructure that makes omitted and undersized colonial territories visible. The technical affordances of visualization function as epistemological arguments, with alluvial diagrams contextualizing decolonization within broader regime transitions, resisting its framing as a discrete historical rupture. Expandable typologies expose how methodological aggregation obscures colonial complexity. The interface is designed to generate boundary objects that serve distinct epistemic communities, supporting cross-disciplinary dialogue without enforcing interpretive consensus. This work contributes methodologically to decolonial digital humanities by arguing that visualization design choices actively shape the analytical possibilities available to users. To counter the treatment of colonial data as neutral input, we argue that interface decisions can be used to embed critical reflection into the analytical process, creating space for users to interrogate instead of passively accept predetermined colonial narratives, and inviting the kind of sustained interpretive engagement that critical scholarship requires.

\backmatter

\bmhead{Acknowledgements}
Varieties of Political Regimes was financed from 2022 to April 2026 by the Saxon State government out of the State budget approved by the Saxon State Parliament.
Christofer Meinecke acknowledges the financial support by the Federal Ministry of Education and Research of Germany and by Sächsische Staatsministerium für Wissenschaft, Kultur und Tourismus in the programme Center of Excellence for AI-research ``Center for Scalable Data Analytics and Artificial Intelligence Dresden/Leipzig``, project identification number: SCADS24B

\bibliography{sn-bibliography}% common bib file

\end{document}